# Denoising and Baseline Correction of Low-Scan FTIR Spectra: A Benchmark of Deep Learning Models Against Traditional Signal Processing

Azadeh Mokari [1,2], Shravan Raghunathan [2], Artem Shydliukh [2], Oleg Ryabchykov [1,2], Christoph Krafft [1,2] and Thomas Bocklitz [1,2], *

[1] Leibniz Institute of Photonic Technology, Member of Research Alliance "Leibniz Health Technologies",07745 Jena, Germany
[2] Institute of Physical Chemistry, Friedrich Schiller University Jena, 07743 Jena, Germany
* Correspondence: thomas.bocklitz@uni-jena.de

**Abstract**

High-quality Fourier Transform Infrared (FTIR) imaging usually needs extensive signal averaging to reduce noise and drift which severely limits clinical speed. Deep learning can accelerate imaging by reconstructing spectra from rapid, single-scan inputs. However, separating noise and baseline drift simultaneously without ground truth is an ill-posed inverse problem. Standard black-box architectures often rely on statistical approximations that introduce spectral hallucinations or fail to generalize to unstable atmospheric conditions. To solve these issues we propose a physics-informed cascade Unet that separates denoising and baseline correction tasks using a new, deterministic Physics Bridge. This architecture forces the network to separate random noise from chemical signals using an embedded SNIP layer to enforce spectroscopic constraints instead of learning statistical approximations. We benchmarked this approach against a standard single Unet and a traditional Savitzky-Golay/SNIP workflow. We used a dataset of human hypopharyngeal carcinoma cells (FaDu). The cascade model outperformed all other methods, achieving a 51.3% reduction in RMSE compared to raw single-scan inputs, surpassing both the single Unet (40.2%) and the traditional workflow (33.7%). Peak-aware metrics show that the cascade architecture eliminates spectral hallucinations found in standard deep learning. It also preserves peak intensity with much higher fidelity than traditional smoothing. These results show that the cascade Unet is a robust solution for diagnostic-grade FTIR imaging. It enables imaging speeds 32 times faster than current methods.

## 1. Introduction

Fourier Transform Infrared (FTIR) imaging creates label-free, high-resolution biochemical maps. It is a powerful tool for investigating metabolic heterogeneity in cancer cell lines, such as the hypopharyngeal carcinoma (FaDu) model [1-3]. However, the trade-off between acquisition speed and spectral fidelity limits the clinical use of FTIR [4, 5]. This is particularly relevant for small, thin samples such as single cells, which exhibit relatively weak absorption bands. Standard protocols average many scans (e.g., 32 to 128) per pixel to suppress instrumental noise and atmospheric interference. This process makes high-quality (HQ) imaging too slow for large tissue sections. Single-scan acquisition is much faster, but it yields low-quality (LQ) spectra. These inputs are dominated by high-frequency noise and non-linear baseline drifts caused by scattering and water vapor [6, 7]. Therefore, rapid clinical FTIR needs computational methods to recover HQ chemical signatures from noisy, single-scan inputs.

Current restoration strategies are primarily divided into traditional signal processing and data-driven deep learning [8-11]. The standard chemometric workflow combines Savitzky-Golay (SG) smoothing for denoising [12, 13] with the Sensitive Nonlinear Iterative Peak (SNIP) algorithm for baseline correction [14, 15]. Although interpretable, this approach has limitations. SG filters act as low-pass filters that systematically broaden spectral peaks and degrade resolution [16-18], while the iterative nature of SNIP introduces computational latency [19]. In contrast, deep learning architectures (like Unet variants) offer rapid, non-linear reconstruction. However, standard neural networks act as black boxes without explicit physical constraints. Separating noise and background drift without a known ground truth is an ill-posed inverse problem which is common when training on real experimental data. This ambiguity forces the network to use statistical approximations. This leads to two failures: hallucinating spectral features and instability on out-of-distribution data [18].

We propose a physics-informed cascade Unet architecture to combine the robustness of physical models with the power of deep learning. Standard monolithic models try to learn the correction in one step. Unlike them, our architecture separates the task into two stages, linked by a physics-informed bridge.

This bridge integrates a deterministic, non-learnable SNIP layer into the forward pass. This design ensures that we handle baseline suppression using established physical principles. By transferring the burden of modeling background variance and baseline drift to the SNIP algorithm, the neural components are freed to focus exclusively on high-frequency denoising and fine-scale artifact refinement.

We show that this hybrid architecture yields superior peak fidelity and stability compared to traditional workflows and standard, end-to-end deep learning models. This approach validates a pathway for high-speed FTIR analysis. It enables reliable interpretation from extremely low-scan acquisitions (e.g., 1-scan).

## 2. Materials and methods

### 2.1 Dataset description

To evaluate the proposed denoising architectures on complex biological data, we utilized FTIR spectroscopic images derived from the FaDu cells cultivated on 1 mm thick CaF2 windows. FaDu refers to a human hypopharyngeal squamous cell carcinoma cell line commonly used in head and neck oncology research. These samples were selected to ensure the model encounters realistic biochemical heterogeneity, as varying cell densities and metabolic activities within the samples produce diverse spectral signatures rather than uniform synthetic signals.

Data were acquired with the FTIR spectrometer 670 and the microscope 620 (Agilent, USA) equipped with a liquid-nitrogen-cooled 64×64 focal plane array (FPA) detector. Acquisition was performed in transmission mode using a 15× Cassegrain condenser and objective lenses as previously described [20] on four distinct sample regions, designated as FaDu1 through FaDu4. These independent fields of view (FOVs) provide the variance needed for robust cross-validation. This ensures the model generalizes across different biological structures. Notably, the dataset includes real-world experimental variability; for instance, the FaDu3 sample shows spectral drift due to environmental instability (e.g., humidity or purge gas fluctuations). This provides a rigorous test for denoising stability in non-ideal conditions.

To construct a supervised learning benchmark, it is essential to have paired noisy and clean representations of the exact same biological target. Consequently, each FOV was imaged under three distinct signal accumulation settings.

The LQ inputs were acquired using 1 scan and 8 scans per pixel. These modes simulate high-speed imaging scenarios where acquisition speed is prioritized, resulting in data characterized by lower signal-to-noise ratios (SNR) and significant instrumental noise contributions.

The HQ ground truth were acquired using 32 scans per pixel. The 32 scans spectra possess a statistically superior SNR and serve as the high quality ground truth for model training and quantitative evaluation.

The resulting hyperspectral data cubes are three-dimensional tensors. Each image has spatial dimensions of 64×64 pixels. The spectral dimension consists of 1584 data points that covers the mid-infrared fingerprint and functional group regions (approximately 950 to 4000 cm$^{-1}$, with a spectral resolution of 4 cm$^{-1}$ and a data interval of roughly 2 cm$^{-1}$ at interleave factor 1). This high-dimensional structure forces the algorithms to handle both spatial correlations between pixels and complex spectral features in biological tissues. A visual comparison of the signal quality across the different scan accumulations is presented in Figure 1.

### 2.2 Dataset preparation and preprocessing

The intrinsic complexity of FTIR spectra interpretation is frequently impeded by the existence of extraneous factors and measurement variability. Variations occur over time within the same sample or between biological replicates. Common causes include biological drift, sample deterioration, inconsistent preparation, and other uncontrolled experimental factors [21, 22].

Such sources of variation, if not addressed, can obscure meaningful spectral patterns and compromise the reliability of downstream analyses [23, 24].

To address these issues, the raw hyperspectral data cubes underwent a comprehensive two-phase processing pipeline. This pipeline was designed first to isolate valid biological signals from physical artifacts (data preparation), and second, to standardize signal distributions for optimal neural network convergence (data preprocessing) as illustrated in Figure 2.

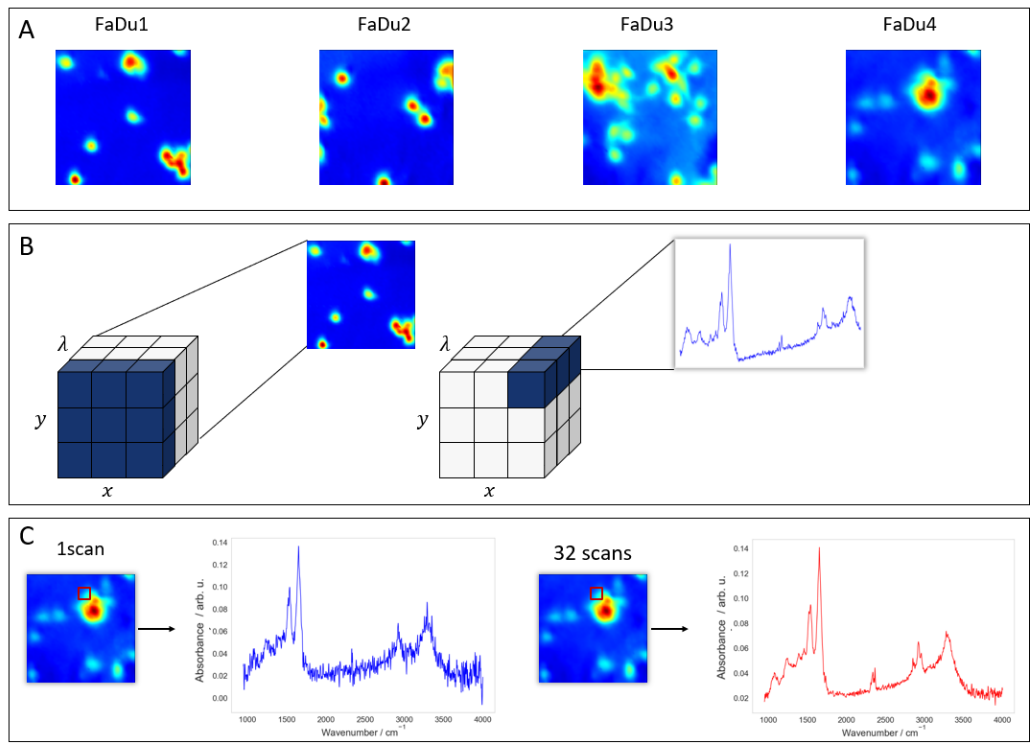

**Figure 1:** (A) Chemical images displaying the four distinct sample regions (FaDu1–FaDu4). (B) Ilustration of the hyperspectral data cube format (x, y, λ). The schematic demonstrates how a slice along the λ-axis represents a spatial map, while each individual spatial pixel corresponds to a full infrared absorbance spectrum containing the sample's biochemical fingerprint. (C) Comparison of spectral quality based on scan accumulation. The panels show spectra extracted from the same spatial pixel (indicated by the red square) using 1 scan versus 32 scans, highlighting the improvement in signal-to-noise ratio with increased scanning.

**2.2.1 Data preparation**

The initial phase focused on extracting cellular spectra from the hyperspectral image cubes. To distinguish biologically relevant information from the empty substrate, a binary mask was generated using Otsu's thresholding method [25]. This thresholding was applied to the integrated intensity of the fingerprint region and the C-H stretching bond region. Consequently, only foreground pixels corresponding to cellular material were extracted for analysis, reducing computational overhead from non-informative background data.

Atmospheric fluctuations during scanning often introduce artifacts. To remove these environmental interferences, specifically water vapor contributions, an average background spectrum was computed from the non-sample pixels and subtracted from each foreground spectrum [26].

Finally, to eliminate regions containing no biological information, the spectral axis was pruned. This involved trimming the noisy detector edges and removing the silent region between 2250 cm⁻¹ and 2401 cm⁻¹, which contains to the strong atmospheric $CO_2$ absorption band.

**2.2.2 Data preprocessing**

Following extraction, the workflow applied differential processing to define the supervised learning task. The goal is to train the model to map raw, distorted inputs to pure-absorbance targets. For the HQ target data (32 scans), the SNIP algorithm was applied immediately to remove the baseline, ensuring the network trained against pure absorbance targets with no baseline offsets. In contrast, the LQ (1 and 8 scans) input spectra bypassed the baseline removal step. Retaining the native baselines forces the model to learn the complex task of simultaneously suppressing the background and reducing random noise.

Following this split, we applied a two-step normalization process to both datasets to ensure numerical stability. First, Standard Normal Variate (SNV) normalization [27] was applied individually to each spectrum to reduce physical scattering effects. Second, a global min-max scaling was performed to map all spectral intensities into a stable [0, 1] range based on the global minimum and maximum values of the dataset.

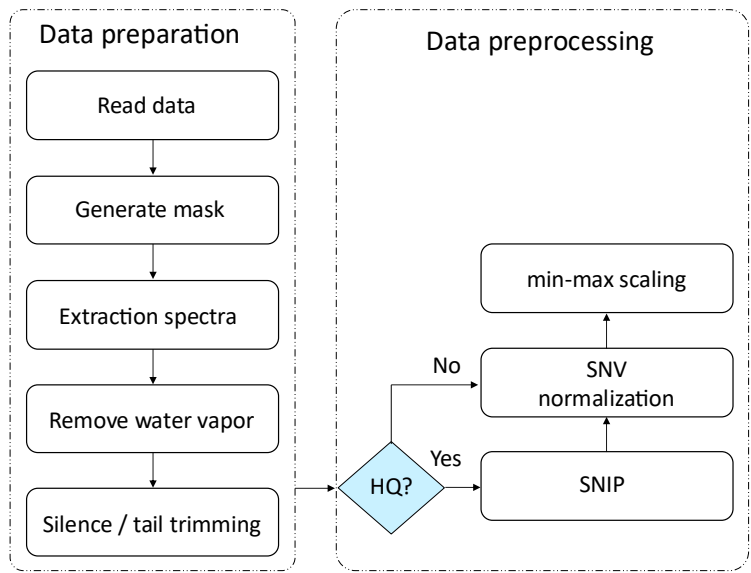

**Figure 2:** Schematic of the computational workflow illustrating the separation between data preparation and data preprocessing. The preparation phase isolates valid biological spectra through masking, water vapor correction, and spectral trimming. The preprocessing phase applies SNIP only to the HQ ground truth, while LQ inputs retain their baselines, before both undergo identical SNV and min-max normalization.

### 2.3 Architectures implemented

#### 2.3.1 Method a: Traditional signal processing (benchmark)

To establish a rigorous baseline for evaluating deep learning performance, we implemented a classic chemometric workflow that sequentially addresses the two primary sources of spectral corruption: high-frequency instrumental noise and low-frequency baseline drift. This pipeline combines SG smoothing with the physics-based SNIP algorithm. Unlike data-driven models that approximate functions via learned weights, this method relies strictly on deterministic mathematical transformations based on signal processing theory.

The initial processing stage addressed high-frequency instrumental noise within the LQ input spectra using a Savitzky-Golay filter. While effective, the SG filter is traditionally sensitive to hyperparameter selection, where an insufficient window length fails to remove noise and an excessive length induces signal distortion [28]. To eliminate arbitrary manual tuning and ensure the benchmark represented the theoretical limit of traditional processing, we utilized the Optuna framework to perform an automated Bayesian hyperparameter search [29]. Using the Tree-structured Parzen Estimator (TPE) algorithm, we optimized the filter's window length and polynomial order to minimize the (Mean Square Error) MSE between the smoothed LQ input and the HQ ground truth, ensuring the method was tuned specifically to the noise characteristics of the FaDu dataset.

A key challenge in this workflow is the domain incompatibility between denoising and baseline correction. While denoising was optimized in the normalized [0,1] domain to maintain metric consistency with the neural networks, the subsequent baseline correction algorithm requires physical intensity values to function correctly. The SNIP algorithm operates on geometric principles, specifically the iterative comparison of local intensity minima, which depend on the relative amplitude ratios of the original absorbance data. Normalization distorts these ratios, making geometric clipping ineffective.

To address this, we introduced a distinct architectural component termed the "Physics Bridge". Mathematically, this is an inverse normalization (reversing the global min-max scaling and SNV normalization). However, it serves a critical function and a necessary transition from the statistical feature space (optimized for gradient descent and filtering) to the physical signal space (absorbance). This restoration ensures that the subsequent baseline correction is driven by spectroscopic constraints rather than statistical artifacts. Once restored to the physical domain, the spectra were processed using the SNIP algorithm to estimate and subtract the background continuum, effectively treating the broad underlying baseline as a low-frequency carrier wave to be isolated from the high-frequency chemical signal.

### 2.3.2 Method b: single Unet

To represent current data-driven methods, we implemented a standard single Unet architecture as an end-to-end regression model [30]. Unlike the traditional workflow, which decouples signal corruption into distinct frequency domains (treating noise and baseline as separate problems), this architecture assumes it can learn a direct non-linear mapping from the raw LQ input to the corrected HQ target. This makes optimization difficult. The network must learn to suppress high-frequency stochastic noise while simultaneously identifying and subtracting low-frequency baseline drifts solely based on learned statistical correlations from the training examples.

Structurally, the model is designed as a 1D convolutional encoder-decoder network with mechanisms to improve feature selectivity. The encoder path utilizes successive convolutional blocks to downsample the input, extracting high-level latent features essential for distinguishing signal from noise. To mitigate the vanishing gradient problem and improve information flow through the deep network, the bottleneck section incorporates residual (ResNet) blocks [31]. Crucially, the decoder path is reinforced with spectral attention modules, which allow the network to dynamically recalibrate feature weights, thereby focusing capacity on chemically relevant spectral bands while suppressing irrelevant background signals.

To preserve high-frequency spatial information lost during downsampling, the network employs skip connections that concatenate features from the encoder directly to the decoder. In the context of vibrational spectroscopy, these connections are critical for retaining the structural integrity of sharp absorbance peaks that might otherwise be eroded by pooling operations. The model was trained in a fully supervised manner, mapping normalized LQ spectra (containing native baselines) directly to normalized, SNIP-corrected HQ targets. However, this black box approach operates without explicit physical constraints. By forcing the model to statistically approximate the geometric process of baseline removal, it becomes susceptible to spectral hallucinations, the generation of artifacts that mimic chemical features, particularly when inferencing on samples with environmental conditions that deviate from the training distribution.

### 2.3.3 Method c: cascade Unet (physics-informed architecture)

To overcome the limitations of both rigid signal processing and black-box deep learning, we developed the physics-informed cascade Unet. This hybrid architecture separates spectral restoration task into two learnable phases, denoising and refinement, separated by a deterministic Physics Bridge. Unlike the single Unet, which must approximate the complex transformation from noisy, tilted inputs to flat, clean outputs in a single pass, the cascade model distributes these tasks to specialize the network components.

The first stage comprises a Unet trained strictly for noise suppression. Its input is the normalized LQ spectrum, containing both high-frequency noise and variable baseline drift. The target of optimization for this stage is the HQ spectrum with the baseline retained. This target definition prevents the network from having to distinguish low-frequency background drift from broad chemical features. This ambiguity is a primary source of hallucination in standard models. Consequently, Unet 1 acts as a high-fidelity denoiser. It learns to suppress stochastic noise while preserving the underlying spectral topology and background curvature.

The Physics Bridge sits between the two neural stages. It is a non-learnable computational layer that enforces spectroscopic constraints. The SNIP algorithm relies on geometric peak clipping that depends on absolute intensity ratios. Therefore, it does not function correctly in the normalized [0, 1] feature space. To address this, the bridge performs a differentiable inversion within the model graph. The output of Unet 1 is first mathematically inverted (inverse min-max followed by inverse SNV) using the input's stored statistics to restore physical absorbance units. Subsequently, the restored spectra are processed by a TensorFlow-wrapped SNIP layer. This ensures that background removal follows established physical laws rather than statistical approximations. It provides stability against environmental drifts that typically confuse standard neural networks.

The baseline-free spectra emerging from the bridge are re-normalized and passed to the second stage, Unet 2. While SNIP is robust, it remains a mathematical approximation that can introduce systematic artifacts, such as the attenuation of broad peak bases or residual curvature. Unet 2 acts as a refinement network trained against the final baseline-free HQ ground truth. It learns a residual correction function to repair SNIP artifacts and recover high-resolution peak shapes. The architecture is trained using a deep supervision strategy with a multi-term loss function, computing the MSE at two distinct exit points: $\text{Loss}_{stage1}$ compares the output of Unet 1 (denoised + baseline) against the HQ target, while $\text{Loss}_{stage2}$ compares the final output of Unet 2 against the baseline-free HQ target. The complete workflow, contrasting the multi-objective training path with the streamlined inference pipeline used for processing new clinical samples, is illustrated in Figure 3.

During the inference (test) phase, the optimization pathways, including the HQ targets and loss calculation blocks shown in Figure 3A, are deactivated. The deployed model operates as a feed-forward pipeline (Figure 3B); a new LQ spectrum is denoised by Unet 1, processed through the deterministic Physics Bridge to remove the baseline, and finally refined by Unet 2. Unlike standard domain adaptation techniques that might be removed after training, the Physics Bridge remains an active, integral component of the inference graph. This ensures that every prediction, whether in training or clinical deployment, adheres to the same spectroscopic constraints.

### 2.4 Experimental design

#### 2.4.1 Cross-validation strategy

To ensure the clinical applicability and robustness of our models, we used a strict Leave-one-sample-out cross-validation (LOSO-CV) strategy [32]. The dataset was partitioned by biological sample, corresponding to the four independent fields of view labeled FaDu 1 through FaDu 4. In each validation iteration, models were trained on spectra from three sample regions and evaluated exclusively on the held-out fourth sample. This design was chosen specifically to prevent data leakage, ensuring that reported metrics reflect the model's ability to generalize to unseen biological heterogeneity rather than its capacity to memorize sample-specific spectral distributions.

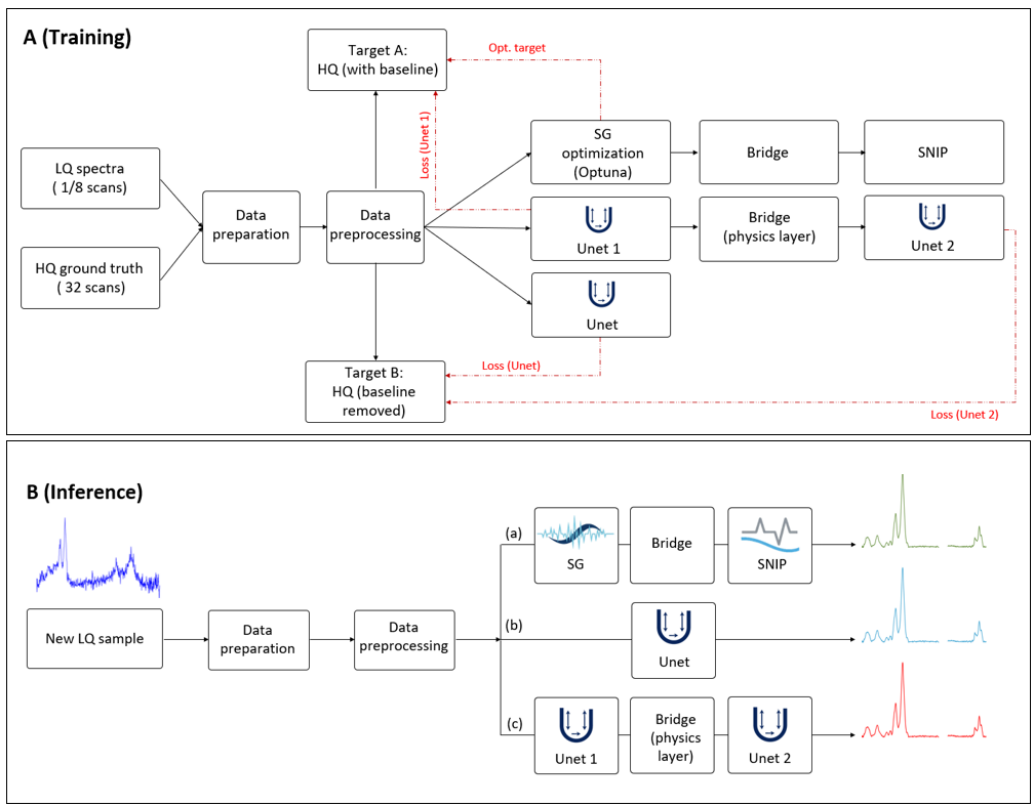

**Figure 3:** Computational workflow comparing multi-objective training and feed-forward inference. (A) Training phase: the diagram shows the parallel implementation of the three benchmarking methods. LQ inputs (1 & 8 scans) and HQ ground truth (32 scans) undergo identical data preparation and preprocessing. The proposed cascade Unet uses a decoupled optimization strategy. Unet 1 minimizes $\text{Loss}_{stage1}$ against Target A (HQ spectra with native baseline), focusing on denoising. The output is then processed by the Physics Bridge, which performs domain inversion and deterministic SNIP baseline removal, before passing to Unet 2, which minimizes $\text{Loss}_{stage2}$ against Target B (baseline-free HQ spectra). Red dashed lines indicate gradient and loss calculations. (B) Inference phase: the deployment workflow for processing unseen clinical samples. Optimization pathways and ground truth targets are deactivated. The model operates as a strictly feed-forward pipeline where the Physics Bridge remains active, ensuring baseline removal follows geometric spectroscopic constraints rather than learned approximations.

**2.4.2 Hyperparameter optimization**

For the traditional benchmark (Method a), we optimized signal processing parameters instead of using arbitrary defaults ensuring a fair comparison. We utilized the Optuna framework to conduct an automated Bayesian hyperparameter search using the Tree-structured Parzen Estimator (TPE) algorithm. We optimized the Savitzky-Golay filter's window length and polynomial order to minimize the (Root Mean Square Error) RMSE between the processed LQ input and the HQ ground truth. This optimization protocol ensures that the benchmark represents the theoretical limit of traditional smoothing capabilities for the specific noise characteristics of this dataset.

**2.4.3 Neural network training protocol**

Both deep learning architectures, the single Unet and cascade Unet, were implemented in TensorFlow and trained using identical hyperparameters to isolate the impact of architectural differences. Training utilized the Adadelta optimizer with a learning rate of 0.05 and a batch size of 32. To prevent overfitting, we used an early stopping mechanism with a patience of 30 epochs. Models were trained for a maximum of 5000 epochs, with the best-performing weights restored for final inference.

**2.4.4 Evaluation framework**

Model performance was assessed using a comprehensive suite of metrics designed to capture both statistical accuracy and chemical validity. To ensure a fair comparison across methods with differing output domains, all metrics were calculated against the normalized, baseline-free HQ ground truth.

To quantify general reconstruction fidelity, we calculated the RMSE and MAE (Mean Absolute Error). However, as Euclidean distance metrics can be insensitive to shape preservation, we essentially utilized the Spectral Angle Mapper (SAM) to measure vector similarity, alongside the Pearson Correlation Coefficient (PCC) and Spearman's Rank Correlation to quantify spectral shape consistency.

Because global error metrics can often mask localized distortions, we implemented targeted chemical fidelity checks. Peak Position Error quantified the wavenumber shifts (cm$^{-1}$) of diagnostic bands, serving as a proxy for chemical identification accuracy. Relative Intensity Bias measured systematic over- or under-estimation of absorbance, which is critical for quantitative concentration analysis.

## 3. Results

**3.1 Quantitative performance (global metrics)**

The quantitative evaluation of spectral reconstruction accuracy reveals a distinct performance hierarchy among the tested architectures, with the physics-informed approach demonstrating superior fidelity across all global metrics. To establish a standardized baseline for improvement, we calculated the error of the raw LQ inputs relative to the HQ ground truth. As summarized in Table 1, the traditional SG+SNIP workflow achieved a baseline reduction in RMSE of 33.67% compared to the raw input. The standard single Unet improved upon this benchmark, yielding a 40.23% reduction in RMSE, validating the capability of deep learning to extract signal from noise more effectively than fixed mathematical filters. However, the proposed cascade Unet significantly outperformed both comparators, achieving an RMSE reduction of 51.30%. This trend was consistent across MAE metric, where the cascade model demonstrated a 52.13% error reduction, compared to 42.34% for the single Unet and 34.66% for the traditional workflow.

Table 1: Quantitative comparison of global spectral reconstruction fidelity. The table summarizes the percentage error reduction for each architecture relative to the raw LQ baseline. The cascade Unet outperformed the other methods across all metrics (RMSE, MAE, and SAM). It achieved reductions of 51.30%, 52.13%, and 51.52% respectively, outperforming both the traditional SG + SNIP workflow and the standard single Unet. Bold values indicate the best performance in each category.

| Method | RMSE reduction (%) | MAE reduction (%) | SAM reduction (%) |
|---|---|---|---|
| **cascade** | **51.30** | **52.13** | **51.52** |
| **SG + SNIP** | 33.67 | 34.66 | 33.71 |
| **single Unet** | 40.23 | 42.34 | 40.97 |

Beyond intensity reconstruction, we assessed the preservation of spectral shape using the SAM and PCC. The raw LQ spectra showed a baseline PCC of 0.87, indicating significant corruption of the biochemical fingerprint. While all three denoising methods successfully restored the correlation to above 0.99, distinguishing their performance required the more sensitive vector-based SAM metric. The cascade Unet reduced the spectral angle error by 51.52%, providing the closest vector alignment to the ground truth. In comparison, the single Unet and SG+SNIP methods achieved reductions of 40.97% and 33.71%, respectively. These results, visualized in Figure 4, confirm that the cascade architecture yields the most chemically accurate spectral shapes.

### 3.2 Stability and robustness

While global averages demonstrate the superiority of the cascade architecture, an analysis of sample-specific performance reveals a critical distinction in model stability. Real-world spectroscopic data is frequently compromised by environmental instabilities, such as fluctuations in purge gas composition or humidity levels, which induce complex, non-linear baseline drifts. We quantified acquisition stability using silent-region noise, defined as the standard deviation of absorbance in the $CO_2$ silent window (2250–2401 cm$^{-1}$). Under predominantly random noise, scan averaging should reduce this metric approximately as $1/\sqrt{N}$ (expected ratios: 1→8 ≈ 2.8, 1→32 ≈ 5.7). FaDu3 deviated strongly from this behavior (1→8: 1.81, 1→32: 3.68) and retained the highest residual silent-region noise at 8 and 32 scans ($1.35 \times 10^{-4}$ and $6.64 \times 10^{-5}$), consistent with drift-like, non-averageable variability.

Consequently, this dataset served as a rigorous stress test for evaluating model robustness against non-ideal acquisition conditions.

Under these challenging conditions, the standard single Unet exhibited marked instability. As illustrated by the sample-wise RMSE boxplots (Figure 5), this architecture produced a wide distribution of errors, notably failing to maintain consistency on the drift-affected sample. This failure mode stems from the black-box nature of the architecture. Without explicit physical baseline knowledge, the model interpreted environmental drift as biological signal. This resulted in the generation of spectral hallucinations where the model effectively encoded the background noise into its prediction.

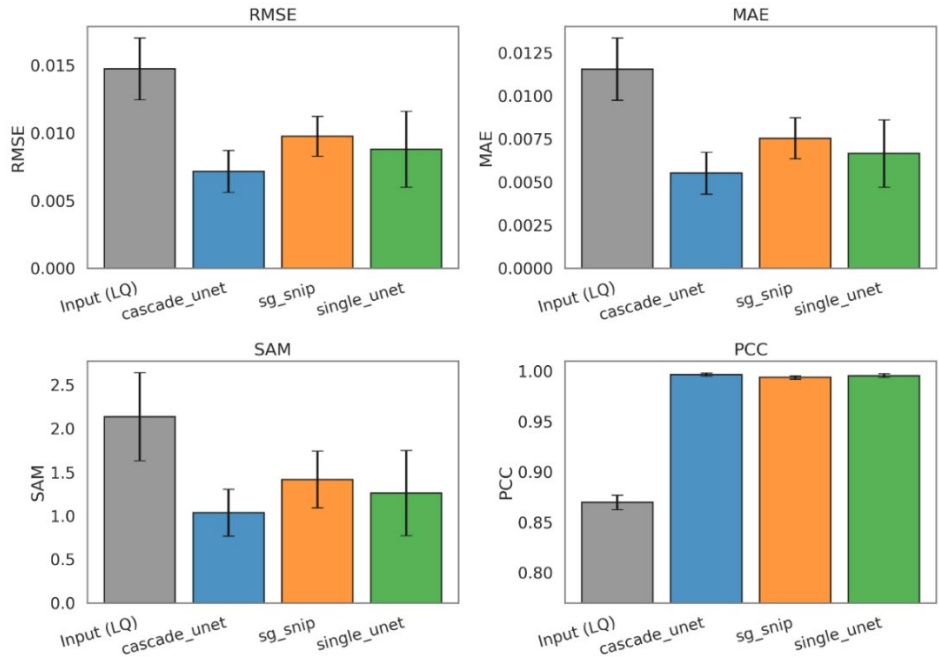

**Figure 4:** Comparative analysis of global spectral reconstruction fidelity across benchmarking methods. The bar charts display the mean performance metrics calculated against the HQ ground truth for the raw LQ input (gray), the proposed cascade Unet (blue), the traditional SG + SNIP workflow (orange), and the standard single Unet (green). Error bars represent the standard deviation across the LOSO-CV. (Top Row) Intensity-based metrics: RMSE and MAE. Lower values indicate better performance. The cascade Unet achieves the lowest error in both categories. (Bottom Row) Shape-based metrics: SAM, where lower values indicate better vector alignment, and PCC, where values closer to 1.0 indicate higher shape similarity. The cascade Unet demonstrates superior preservation of spectral topology compared to both the traditional and single-stage deep learning benchmarks.

In contrast, the cascade Unet showed exceptional resilience to these environmental changes. By offloading the baseline removal task to the deterministic physics bridge (the embedded SNIP layer), the model became immune to background drift. The network did not have to learn the stochastic properties of environmental drift. This allowed the optimization to focus exclusively on signal restoration. This stability is visually confirmed by the plots in Figure 5, where the cascade model maintains a consistent, symmetric performance profile across all four biological samples. Conversely, the single Unet displays significant skewing and degradation in the presence of drift. The physics-informed approach thus delivers not only lower absolute error but also the reliability requisite for clinical pathology, where environmental conditions cannot always be perfectly rigorously controlled.

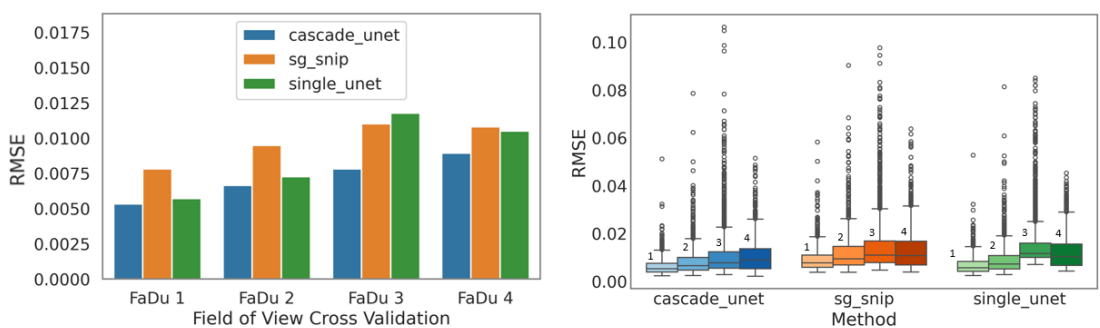

**Figure 5:** Assessment of model stability and generalization robustness across variable environmental conditions. (Left) RMSE stratified by LOSO-CV (FaDu 1–4). The cascade Unet (blue) maintains consistent low-error performance across all biological samples. In contrast, the single Unet (green) and traditional SG+SNIP (orange) exhibit notable performance degradation on the FaDu 3 dataset, which is characterized by significant environmental drift. (Right) Boxplots illustrating the distribution of reconstruction errors for each method. The single Unet displays a wider interquartile range and a high density of outliers, indicating instability and the generation of spectral artifacts. The cascade Unet achieves the most compact error distribution, confirming that the deterministic Physics Bridge effectively immunizes the model against baseline instabilities.

### 3.3 Peak fidelity analysis

Although global error metrics provide a macroscopic summary of reconstruction fidelity, they fail to explicitly quantify the preservation of diagnostic biochemical features. To validate the model's utility for rigorous chemical analysis, we performed a peak-aware evaluation focusing on two critical parameters of infrared spectroscopy: peak localization and intensity conservation.

#### 3.3.1 Peak position and localization

Precise wavenumber localization is fundamental for the correct identification of specific molecular bonds. Our analysis of absolute peak position error shows that the traditional SG+SNIP workflow produced the largest and most dispersed errors as shown in Figure 6. This inaccuracy is largely attributable to the inherent limitations of polynomial smoothing, which effectively acts as a low-pass filter that can induce phase shifts in peak centers, particularly within asymmetric spectral bands. In contrast, both deep learning architectures demonstrated superior localization accuracy. The cascade Unet achieved the tightest error distribution, indicating that the physics-informed model effectively distinguishes true signal maxima from noise without introducing the spatial distortions common to linear filters.

#### 3.3.2 Peak intensity and quantitative bias

Since peak height serves as a direct proxy for molecular concentration, systematic bias in intensity reconstruction can lead to quantitative errors. Table 2 summarizes the reconstruction metrics for the three methods. The traditional method showed a systematic negative bias, consistently underestimating peak heights by approximately 1.6%. This attenuation effect is characteristic of smoothing algorithms that average peak maxima with surrounding lower-intensity values.

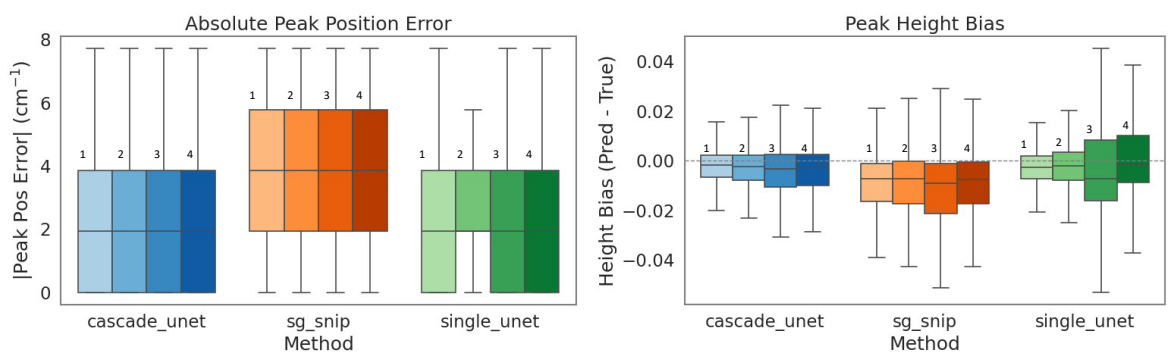

**Figure 6:** Evaluation of diagnostic chemical fidelity through peak-aware metrics. (Left) Absolute Peak Position Error: Boxplots show peak localization accuracy. The traditional SG+SNIP method (orange) has the largest deviations, attributable to phase shifts induced by polynomial smoothing. The deep learning models achieve tighter localization, with the cascade Unet (blue) minimizing spatial error. (Right) Peak Height Bias: Analysis of quantitative reconstruction accuracy. The dashed line at 0.0 represents zero bias (perfect reconstruction). The traditional method displays a systematic negative bias (median ≈ -1.6%), confirming the attenuation of peak intensity. While the single Unet (green) achieves a low median bias, it suffers from high variance (large interquartile range), indicating instability. The cascade Unet demonstrates superior precision (tightest IQR of 0.024) and minimal bias.

**Table 2**: Comparison of peak intensity reconstruction bias and precision. The table summarizes the distribution of the peak intensity error (Reconstructed - True) for the three denoising methods. Bias is shown by the 25th, 25th (median), and 75th percentiles, while precision is represented by the Interquartile Range (IQR). The cascade Unet achieves the highest precision (lowest IQR of 0.024) and lowest median bias (-0.006 or -0.6%) compared to the traditional SG + SNIP method (median bias of -0.016 or -1.6%) and the single Unet method (highest IQR of 0.039 and median bias of -0.007 or -0.7%). The cascade's superior performance is attributed to the effective baseline removal via the Physics Bridge.

| Method | 25$^{\text{th}}$ perc | Median | 75$^{\text{th}}$ perc | IQR |
|---|---|---|---|---|
| cascade | -0.020 | -0.006 | 0.005 | 0.024 |
| SG + SNIP | -0.032 | -0.016 | -0.002 | 0.030 |
| single Unet | -0.028 | -0.007 | 0.011 | 0.039 |

The single Unet presented a contrasting failure mode; while it achieved a low median bias of -0.7%, it suffered from poor precision, evidenced by a high Interquartile Range (IQR) of 0.039. This high variance indicates that while the single Unet is accurate on average, its predictions fluctuate unpredictably for individual spectra, struggling to determine the correct amplitude in the presence of noise. The cascade Unet emerged as the most reliable estimator, combining the lowest bias (-0.6%) with the highest precision (IQR of 0.024). By removing the baseline via the

Physics Bridge, the cascade network avoids the ambiguity of determining peak amplitude against a shifting background, resulting in tightly clustered and chemically reliable intensity predictions.

### 3.4 Visual inspection outcomes

To complement the statistical performance metrics, we conducted a qualitative visual assessment of the reconstructed spectra to verify the preservation of chemically diagnostic morphology. Figure 7 presents a representative spectral comparison between the three denoising architectures and the HQ ground truth. While all methods successfully recover the overall spectral shape across the fingerprint and functional group regions (1000–3000 cm$^{-1}$), a closer examination of the high-frequency C-H stretching region (2800–3000 cm$^{-1}$) reveals distinct differences in fine-structure preservation.

The traditional SG+SNIP workflow (orange trace) shows the typical limitations of polynomial smoothing filters. As evidenced in the zoomed inset, this method systematically erodes sharp spectral features, resulting in broadened peaks and a noticeable reduction in peak maxima relative to the HQ target. This flattening effect confirms the negative intensity bias quantified in the earlier peak-aware analysis, obscuring subtle shoulder features critical for lipid and protein discrimination.

The single Unet (green trace), while sharper than the traditional method, demonstrates a lack of localized precision. In the C-H stretching region, the model struggles to accurately resolve the valley depth between adjacent peaks and fails to match the absolute absorbance intensity of the ground truth. This visual deviation aligns with the higher interquartile range observed in the quantitative metrics, suggesting that the single-stage model has difficulty converging on the precise amplitude of complex multiplets.

In distinct contrast, the cascade Unet (blue trace) demonstrates a remarkable overlap with the HQ ground truth (black trace). The physics-informed architecture faithfully reproduces the intricate spectral topology, including the exact amplitude of peak maxima and the inflection points of the peak shoulders. By architecturally separating noise suppression from baseline management, the cascade model avoids both the over-smoothing of the Savitzky-Golay filter and the amplitude instability of the single Unet, yielding a reconstruction that is visually indistinguishable from the 32-scan reference.

### 3.5 Computational efficiency

Deploying denoising models in clinical pathology depends on both reconstruction quality and computational efficiency. We evaluated the resource requirements for each method, quantifying learnable parameter counts, training duration, and inference latency per spectrum. As summarized in Table 3, the traditional SG+SNIP workflow operates with zero learnable parameters, as it relies on fixed mathematical kernels rather than weighted neural connections.

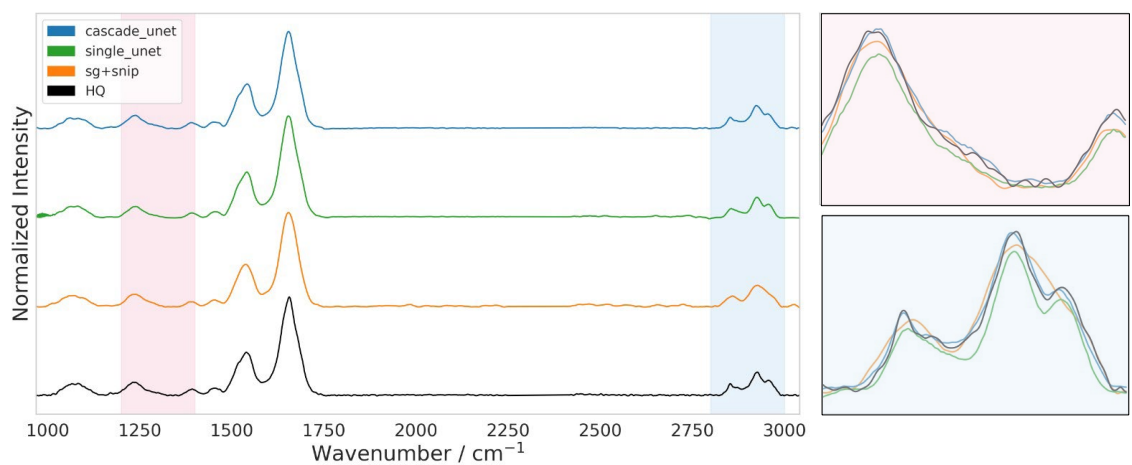

**Figure 7:** Qualitative comparison of spectral reconstruction fidelity in the fingerprint and high-wavenumber regions. Full-range stacked spectra (1000–3050 cm$^{-1}$) for the cascade Unet (blue), single Unet (green), SG+SNIP (orange), and HQ ground truth (black). Shaded areas indicate the two evaluation windows: a representative fingerprint sub-region (pink, 1200–1400 cm$^{-1}$) and the lipid-associated C–H stretching region (blue, 2800–3000 cm$^{-1}$). Right panels show overlaid (no offset) zooms of the highlighted regions to enable direct comparison of peak shapes and amplitudes. SG+SNIP exhibits peak broadening and intensity attenuation, while the single Unet shows deviations in peak shape and amplitude stability. The cascade Unet most closely follows the HQ spectrum, preserving peak height, width, and fine shoulder structures.

Consequently, it requires negligible optimization overhead (0.01 hours), limited only to the automated hyperparameter search (via Optuna) used to define the optimal window length and polynomial order. However, this advantage disappears during inference phase. Despite lacking deep layers, the traditional method requires approximately 0.48 ± 0.02 ms to process a single spectrum. This latency is driven by the iterative nature of the SNIP algorithm and the polynomial calculations of the SG filter, which are typically executed sequentially on the CPU and do not benefit from the massive parallelization capabilities of modern GPU hardware.

In contrast, the deep learning models illustrate the distinct trade-offs inherent in data-driven approaches. The single Unet, comprising approximately 21.5 million parameters, required roughly 3.3 ± 0.3 hours to train across the four validation fields of view. However, it emerged as the most efficient architecture for deployment, achieving an inference speed of just 0.28 ± 0.04 ms per spectrum. This performance, nearly twice the speed of the traditional benchmark, is directly attributable to the highly parallel nature of convolutional operations on the GPU, which allows for the simultaneous processing of large spectral batches.

The proposed cascade Unet, being a dual-stage architecture, represents the most computationally intensive model. It contains approximately 43 million parameters and required an average training duration of 14.8 ± 1.9 hours. Its inference time increased to 2.29 ± 0.57 ms per spectrum. This increased latency is caused by the Physics Bridge bottleneck, where data must be transferred from the GPU to the CPU for the non-differentiable SNIP calculation and subsequently returned to the GPU for the second refinement stage. While significantly slower than the single Unet, this processing speed remains well within acceptable limits for high-throughput imaging. It translates to the processing of a typical 64 ×64 pixel image tile in under 10 seconds, a negligible duration compared to the substantial physical acquisition time saved by scanning at 1 scan rather than 32 scans.

Table 3: Comparative analysis of computational resource requirements and inference latency. The table details the learnable parameter count, total training or optimization duration, and the mean inference time per spectrum for each denoising architecture. Training times represent the cumulative duration across LOSO-CV framework. Inference times are reported as mean ± standard deviation. Note: The traditional SG + SNIP method utilizes zero learnable weights; the reported 0.01 hours corresponds to the hyperparameter optimization phase (window length/polynomial order selection) via Optuna. While the single Unet offers the fastest inference speed due to GPU parallelization, the cascade Unet remains within clinically viable limits for high-throughput imaging despite the computational overhead of the Physics Bridge.

| Method | Parameters | Training time | Test time |
| --- | --- | --- | --- |
| cascade | 42,969,994 | 14.8 ± 1.9 hours | 2.29 ± 0.57 ms |
| SG + SNIP | N/A | 0.01 hours | 0.48 ± 0.02 ms |
| single Unet | 21,484,997 | 3.3 ± 0.3 hours | 0.28 ± 0.04 ms |

## 4. Discussion

The superiority of the cascade Unet can be directly attributed to the structural integration of the Physics Bridge, which fundamentally alters the learning dynamics of the neural network. Standard deep learning models, such as the single Unet, are tasked with learning a complex, high-dimensional mapping that requires the simultaneous suppression of random noise and the estimation of variable background baselines. Our results from the drift-affected FaDu 3 dataset demonstrate the inherent risk of this coupled formulation; when the model encounters environmental conditions or baseline curvatures that deviate from the training distribution, it lacks the physical context to distinguish these anomalies from biological signals, leading to the generation of spectral hallucinations and false positive peaks.

The cascade architecture solves this by separating the optimization tasks. By defining the target for the first stage as the baseline-present spectrum, the first Unet is relieved of the ambiguity associated with background removal, allowing it to act as an efficient, non-linear denoiser. The subsequent Physics Bridge, specifically the deterministic SNIP layer, acts as a non-learnable guardrail. Because SNIP operates on physical intensity values according to rigid geometric constraints, it is immune to the overfitting phenomena that plague learnable parameters. It reliably removes the baseline, even if the specific drift pattern was not in the training set. Consequently, the second stage receives a consistently flattened spectrum, requiring it to correct only minor mathematical artifacts rather than interpret complex environmental interferences.

This architectural choice ensures that the model remains robust to the instrumental variations inevitable in clinical settings.

This study highlights the distinct trade-offs between data-driven and model-driven signal processing. The traditional Savitzky-Golay workflow represents a model-driven approach constrained by fixed mathematical assumptions. While robust and interpretable, it is mathematically rigid; it cannot differentiate between high-frequency noise and sharp chemical features if their frequency components overlap. This limitation was empirically verified by our peak fidelity analysis, where traditional processing systematically broadened peaks and underestimated intensities, effectively reducing the spectral resolution of the imaging system.

Conversely, the single Unet represents a pure data-driven approach. It offers exceptional flexibility and can learn to separate signal from noise in regimes where linear filters fail. However, this flexibility comes at the cost of stability; without physical constraints, the model is prone to overfitting noise patterns and memorizing baselines, leading to poor generalization on unseen or unstable data. The cascade Unet effectively synthesizes these approaches. It uses the learnable power of deep learning to perform superior denoising without resolution loss, while simultaneously anchoring the output to physical reality through the deterministic SNIP baseline correction. This hybrid approach delivers the precision of AI with the reliability of classical spectroscopy.

## 5. Conclusion

The clinical imperative driving this research is the critical need for acquisition speed in spectroscopic pathology. Standard FTIR imaging protocols, which typically require averaging 32 scans per pixel to achieve sufficient signal-to-noise ratios, are prohibitively slow for the routine histopathology of large tissue sections, which requires collecting multiple images and reconstructing them into a mosaic. By validating that single-scan spectra can be computationally restored to a quality statistically indistinguishable from 32 scans ground truth, our findings unlock a potential 32-fold acceleration in data acquisition.

While the cascade Unet introduces a computational overhead, increasing the inference time to approximately 2.29 ms per spectrum compared to 0.28 ms for the single Unet, this latency is negligible in the context of the overall clinical workflow. The time saved by reducing the physical scan duration from hours to minutes vastly outweighs the milliseconds required for computational inference. Furthermore, the rigorous preservation of peak ratios and linewidths ensures that this speed increase does not come at the expense of diagnostic accuracy. Ultimately, this method paves the way for FTIR imaging to transition from a niche research technique to a viable high-throughput modality for clinical diagnostics.

**Author Contributions:** Conceptualization, A.M. and T.B.; methodology, A.M., O. R. and T.B.; software, A.M.; validation, A.M. and T.B.; formal analysis, A.M.; investigation, A.M.; resources, C.K.; data curation, S.R., A.S. and C.K.; writing—original draft preparation, A.M.; writing—review and editing, T.B. and C.K.; visualization, A.M.; supervision, T.B. and C.K.; project administration, T.B. and C.K.; funding acquisition, T.B. and C.K. All authors have read and agreed to the published version of the manuscript.

**Acknowledgments:** We authors thank Nadia Baumann (University Hospital Jena, Germany) for preparation of Fadu cells. This work is supported by BMFTR (Federal Ministry of Research, Technology and Space) within the funding programs QUANCER (13N16444), and funding program Photonics Research Germany (13N15466 (LPI-BT1-FSU), 13N15704 (LPI-BT2-IPHT), 13N15708 (LPI-BT3-IPHT)) and is integrated into the Leibniz Center for Photonics in Infection Research (LPI). The LPI initiated by Leibniz-IPHT, Leibniz-HKI, Friedrich Schiller University Jena and Jena University Hospital is part of the BMFTR national roadmap for research infrastructures.